\documentclass[sigconf,nonacm]{acmart}

\AtBeginDocument{%
  }

\usepackage[hang,small,bf,tight]{subfigure}
\usepackage{listings}

\usepackage[table]{xcolor}
\definecolor{lightgray}{rgb}{0.95, 0.95, 0.95}

\usepackage{enumitem}

\begin{document}

\title{Document Data Matching for Blockchain-Supported Real Estate}

\author{Henrique Lin}
\affiliation{%
  \institution{INESC-ID, Instituto Superior Técnico, Universidade de Lisboa}
  \city{Lisbon}
  \country{Portugal}
}
\email{henrique.lin@tecnico.ulisboa.pt}

\author{Tiago Dias}
\affiliation{
  \institution{Unlockit}
  \city{Lisbon}
  \country{Portugal}
}
\email{tiago.dias@unlockit.io}

\author{Miguel Correia}
\affiliation{%
  \institution{INESC-ID, Instituto Superior Técnico, Universidade de Lisboa}
  \city{Lisbon}
  \country{Portugal}}
\email{larst@affiliation.org}

\renewcommand{\shortauthors}{Lin et al.}

\begin{abstract}
The real estate sector remains highly dependent on manual document handling and verification, making processes inefficient and prone to fraud. This work presents a system that integrates optical character recognition (OCR), natural language processing (NLP), and verifiable credentials (VCs) to automate document extraction, verification, and management. The approach standardizes heterogeneous document formats into VCs and applies automated data matching to detect inconsistencies, while the blockchain provides a decentralized trust layer that reinforces transparency and integrity.
A prototype was developed that comprises (i) an OCR-NLP extraction pipeline trained on synthetic datasets, (ii) a backend for credential issuance and management, and (iii) a frontend supporting issuer, holder, and verifier interactions. Experimental results show that the models achieve competitive accuracy across multiple document types and that the end-to-end pipeline reduces verification time while preserving reliability. The proposed framework demonstrates the potential to streamline real estate transactions, strengthen stakeholder trust, and enable scalable, secure digital processes.
\end{abstract}


\maketitle

\section{Introduction}

The real estate sector is an economic cornerstone in many countries, driving employment, investment, and urban development. However, property transactions remain highly dependent on manual document handling and verification, which are time-consuming, error-prone, and vulnerable to fraud. The diversity of formats, from paper to PDFs and scanned images, and the involvement of multiple stakeholders add additional complexity, causing delays, additional costs, and lack of trust between parties. Strict regulatory frameworks, such as the General Data Protection Regulation (GDPR)~\cite{gdpr2016general} and anti-money laundering (AML) legislation, further increase the burden of compliance.  

Traditional solutions, often based on centralized systems and manual interventions, lack the automation and interoperability required to meet these challenges. Recent studies highlight Blockchain \cite{NISTIR8202} and Self-Sovereign Identity (SSI)~\cite{Allen2016} as promising enablers of decentralized trust. Blockchain provides immutability and transparency, while SSI empowers users with Verifiable Credentials (VCs)~\cite{vcdata2023} to prove document authenticity. However, requiring all entities to immediately issue documents as VCs is unrealistic; existing records such as property deeds, tax statements, and energy certificates must first be integrated. A practical solution must therefore bridge legacy document formats and modern credentialing technologies, enabling gradual adoption.  

This work aims to design and implement the \textbf{Real Estate Credentialing System} (hereafter referred to as the \textbf{Credentialing System}), a prototype platform that:
\begin{itemize}[leftmargin=\parindent]
    \item Automates the extraction of key information from real estate documents using OCR and NLP;
    \item Converts heterogeneous document formats into standardized VCs;
    \item Performs automated verification and data matching to detect inconsistencies;
    \item Integrates blockchain as a trust layer to reinforce transparency and integrity.
\end{itemize}

The main challenges addressed are document heterogeneity, extraction accuracy, and interoperability of VCs across platforms.  

The proposed system contributes in three main ways:
\begin{enumerate}[leftmargin=\parindent]
    \item \textbf{System Implementation} – Development of the Credentialing System prototype, combining OCR, a fine-tuned NLP model trained on synthetic datasets, and backend services for VC issuance and management. The frontend demonstrates issuer, holder, and verifier roles in end-to-end credential flows.
    \item \textbf{Experimental Evaluation} – Assessment of three NLP models on synthetic datasets and validation of the entire pipeline against manual verification, showing competitive extraction accuracy and substantial efficiency gains.
    \item \textbf{Applied Relevance} – Integration of blockchain and SSI concepts into real estate workflows in collaboration with industry, ensuring that the solution addresses real-world requirements.
\end{enumerate}

The remainder of this paper is structured as follows: Section~\ref{sec:background} reviews the background and related work. Section~\ref{sec:design} describes the design of the system. Section~\ref{sec:implement} details the implementation. Section~\ref{sec:evaluation} presents the evaluation. Finally, Section~\ref{sec:conclusion} concludes with lessons learned and directions for future research.

\section{Background and Related Work}
\label{sec:background}

\subsection{Blockchain and Self-Sovereign Identity}
Blockchain provides a decentralized append-only ledger where transactions are grouped into cryptographically linked blocks, ensuring tamper-evidence, auditability, and resilience to single points of failure~\cite{NISTIR8202,nakamoto2008bitcoin}. Depending on governance, blockchains may be public, where anyone can participate, or consortium-based, where access is restricted to authorized entities, but in both cases, the main contribution is the elimination of trust in centralized intermediaries.  

Based on these principles, SSI applies decentralization to the management of digital identity~\cite{Allen2016}. Two core components are Decentralized Identifiers (DIDs) and VCs. DIDs are unique, cryptographically verifiable identifiers that can be anchored in ledgers or resolved through external infrastructures such as \verb|did:web| or \verb|did:key|, each approach presenting trade-offs in scalability, persistence, and cost~\cite{W3Cdid}. VCs are digital attestations signed by trusted issuers according to W3C standards, enabling cryptographic verification of claims without direct issuer participation. They support privacy-preserving techniques such as selective disclosure and zero-knowledge proofs, allowing only the necessary attributes to be revealed.  

To promote interoperability across ecosystems, the OpenID for Verifiable Credentials (OIDC4VC) standard defines protocols for credential issuance and presentation using the widely adopted OpenID Connect framework. This bridges SSI with existing identity infrastructures and facilitates practical adoption. In addition, open-source SSI platforms such as \emph{walt.id} and \emph{Hyperledger Aries} provide implementations of DID resolution, credential issuance, and revocation mechanisms, allowing faster prototyping and deployment in real-world scenarios.  

A final element of SSI is credential life cycle management, which addresses the need to update, suspend, or revoke credentials once they are no longer valid. Common strategies include short-lived credentials, issuer-hosted status checks, shared registries, and the use of separate status credentials~\cite{ebsi_revocation_strategies}. Standardized status formats have also been proposed to enable verifiers to confirm credential validity in a scalable and privacy-preserving way~\cite{ebsi_vc_status}.

\subsection{Data Extraction and Reconciliation}
Automating the handling of documents first requires converting unstructured or semi-structured inputs into reliable structured data. OCR plays a central role in this process by transforming scans and photographs into machine-readable text through stages such as image pre-processing, segmentation, recognition, and post-processing~\cite{ocrSurvey}. Early OCR systems relied heavily on hand-crafted features and rule-based heuristics, but modern tools increasingly employ deep learning architectures that improve robustness to noise, complex fonts, and layout variability. Widely adopted open-source and commercial engines, such as Tesseract, EasyOCR, PaddleOCR, and ABBYY FineReader, illustrate the diversity of solutions available.

However, OCR alone only produces plain text, which often lacks the structure needed for downstream processing. NLP techniques are therefore used to transform raw text into meaningful fields. Classical approaches -- tokenization, part-of-speech tagging, and named entity recognition -- have long been applied to extract information such as names, dates, or locations~\cite{eisenstein2019nlp}. More recent advances leverage deep neural networks and transformer architectures capable of learning contextual and layout-aware representations, which are especially relevant for documents where both textual and visual cues matter. Models such as LayoutLMv3 and Donut combine visual embeddings with text to parse semi-structured layouts directly~\cite{huggingface_transformers}, and their performance is typically enhanced by fine-tuning on domain-specific datasets.

Once key information is extracted, data reconciliation (also known as entity resolution or record linkage) ensures consistency and accuracy across documents that may represent the same entity in different ways. This is particularly important in real estate, where identifiers, such as names, addresses, and tax numbers, must be aligned across multiple sources. Common reconciliation techniques include edit distance and $n$-gram overlap for typographical errors, token-based similarity measures such as Jaccard or cosine similarity for multi-word fields, phonetic encodings to capture spelling variants, and embedding-based similarity to leverage semantic representations~\cite{duplicateRecordDetection}. In geographic contexts, reconciliation can also rely on geocoding services to normalize and cross-check location data. Together, extraction and reconciliation form the foundation for converting heterogeneous legacy documents into structured representations suitable for reliable verification and analysis.

\subsection{Related Work}
\label{sec:relatedWork}

Prior studies align with four main themes relevant to this work: 
(A) blockchain-based data extraction and verification; 
(B) blockchain applications in real estate; 
(C) advanced data extraction techniques; and 
(D) verifiable credential (VC) use cases.

\subsubsection*{Blockchain-Based Data Extraction}
Azzam et al.~\cite{egov} introduced SECHash that integrates OCR and blockchain for e-government invoices to ensure transparency and integrity. Similar works, such as Docschain~\cite{docsChain} and solutions by Mthethwa et al.~\cite{hardcopy} and Gaikwad et al.~\cite{academicCert}, combine OCR and smart contracts for academic or certificate verification. Although these systems demonstrate secure blockchain–OCR integration, they remain limited to structured document types and lack standardized credential formats for interoperability.

\subsubsection*{Blockchain in Real Estate}
Blockchain adoption in real estate mainly focuses on registries and digital transactions. Industry platforms such as \textit{Propy}~\cite{propy} and \textit{Ubitquity}~\cite{ubitquity} use blockchain for property title management and tokenized ownership. Academic initiatives, e.g., Luís et al.~\cite{luis2024real}, use self-sovereign identity (SSI) and Polygon ID for privacy-preserving property data sharing. Although these solutions enhance trust and transparency, they typically assume pre-digitized documents and overlook data extraction from legacy sources.

\subsubsection*{Advanced Data Extraction Techniques}
Recent research emphasizes multimodal AI for unstructured documents. Saout et al.~\cite{invoiceOverview} and Mahadevkar et al.~\cite{mahadevkar2024} survey OCR–NLP pipelines using models such as BERT and LayoutLMv3 for layout-aware extraction. These works achieve high performance on invoices and receipts, but lack evaluation on heterogeneous domain-specific records such as those in real estate.

\subsubsection*{Applications of Verifiable Credentials}
VCs enable decentralized, privacy-preserving document verification. The surveys by Mazzocca et al.~\cite{mazzocca2024survey} highlight the growing adoption in healthcare, logistics, and digital identity frameworks such as DIDKit, Hyperledger Aries, and EBSI. However, most systems rely on pre-structured inputs and rarely couple credential issuance with automated document extraction.

\subsubsection*{Gap and Contribution}
The literature shows that blockchain-OCR integration and VC ecosystems are advancing but often remain isolated or limited to narrow domains. This paper extends previous work by:
\begin{itemize}[leftmargin=\parindent]
    \item Targeting heterogeneous real estate documents (citizen cards, energy certificates, property records);
    \item Integrating OCR, fine-tuned NLP, and VCs into a single automated pipeline;
    \item Demonstrating end-to-end extraction and VC issuance to connect raw data with blockchain-based validation.
\end{itemize}

\section{System Design}
\label{sec:design}

This section presents the \textbf{Credentialing System} design.

\subsection{Overview}
The {Credentialing System} supports the digitization and verification of real estate documents by transforming heterogeneous records into standardized VCs.  
It combines a document processing pipeline (OCR + NLP), a credentialing service for issuance and management, and user-facing interfaces for Holders, Issuers, and Verifiers.  

The main external actors are:
\begin{itemize}[leftmargin=\parindent]
    \item \textit{Holder}: Individuals or organizations (e.g., property owners, tenants) who submit documents such as deeds or rental contracts and receive VCs.
    \item \textit{Issuer}: Authorized entities (e.g., notaries, registrars, government agencies) who validate extracted data, issue credentials, and manage revocation.
    \item \textit{Verifier}: Third parties (e.g., banks, landlords, property platforms) who consume credentials to confirm authenticity, issuer signature, and revocation status.
\end{itemize}

\begin{figure}[tb]
  \centering
  \includegraphics[width=\columnwidth]{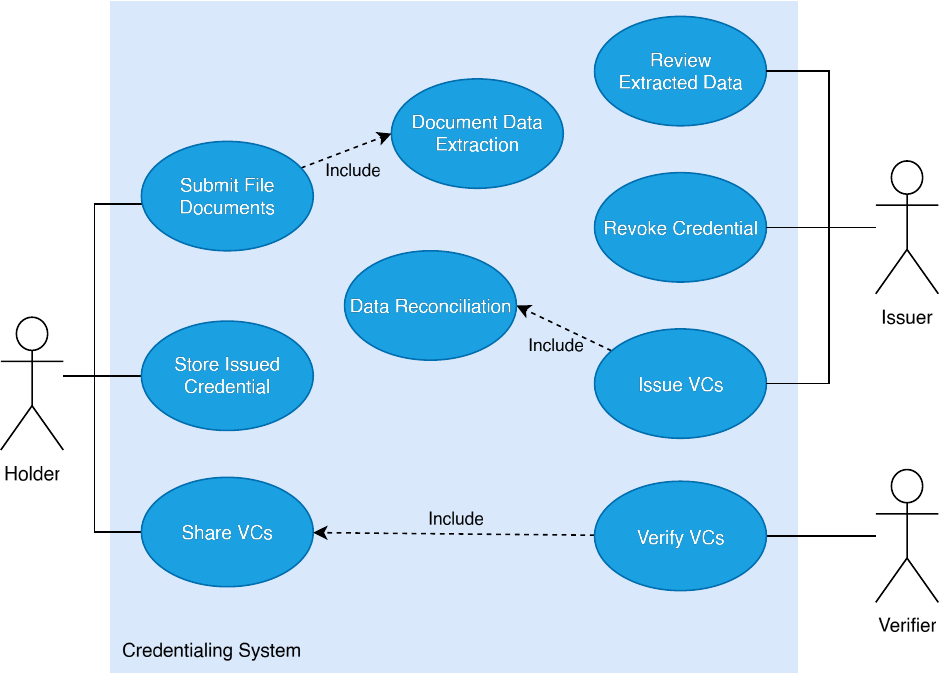}
  \caption{Use Case Diagram for the \textbf{Credentialing System}}
  \label{fig:UseCase}
\end{figure}

Automated steps such as document data extraction and reconciliation run in the background, reducing manual verification effort, and supporting trust in the process.  
Figure~\ref{fig:UseCase} illustrates these actors and their interactions.

\subsection{System Workflows}
The operation of the \textbf{Credentialing System} can be divided into two categories:  
(1) \textit{Credential Issuance}, which covers the end-to-end process of transforming submitted documents into VCs, and  
(2) \textit{Post-Issuance}, which maintains the validity and trustworthiness of credentials through verification and revocation.  

\subsubsection{Credential Issuance Workflow}
\begin{figure}[tb]
  \centering
  \includegraphics[width=\columnwidth]{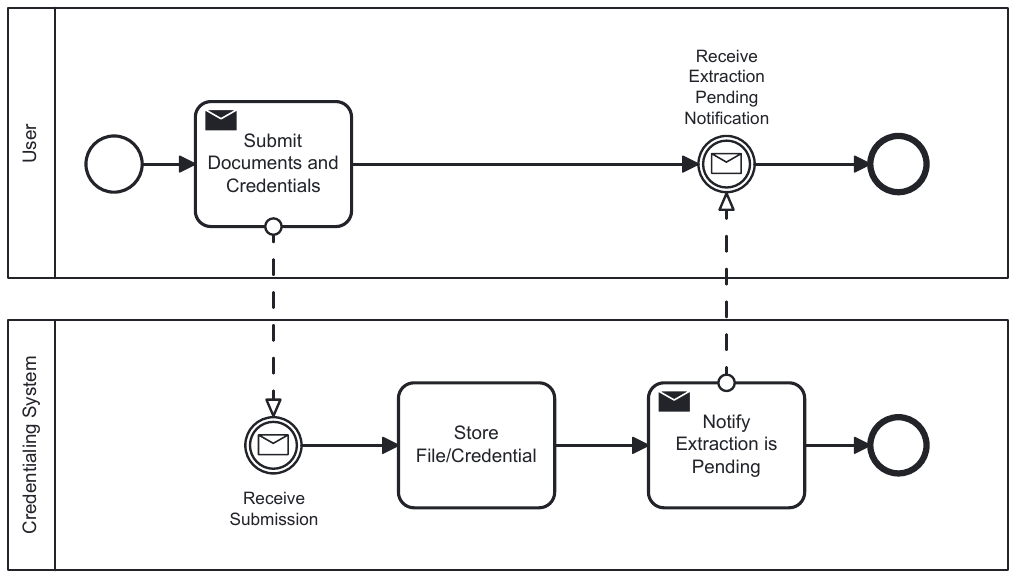}
  \caption{Stage 1 – Data Ingestion Workflow}
  \label{fig:Stage1BPMN}
\end{figure}

The issuance process unfolds in three main stages:  

\paragraph{Stage 1 – Data Ingestion}
The Holder submits the required documents, which may include images, PDFs, or existing VCs. The system stores these inputs in a consistent format to ensure that later steps can operate reliably (Figure~\ref{fig:Stage1BPMN}).  

\begin{figure}[tb]
  \centering
  \includegraphics[width=\columnwidth]{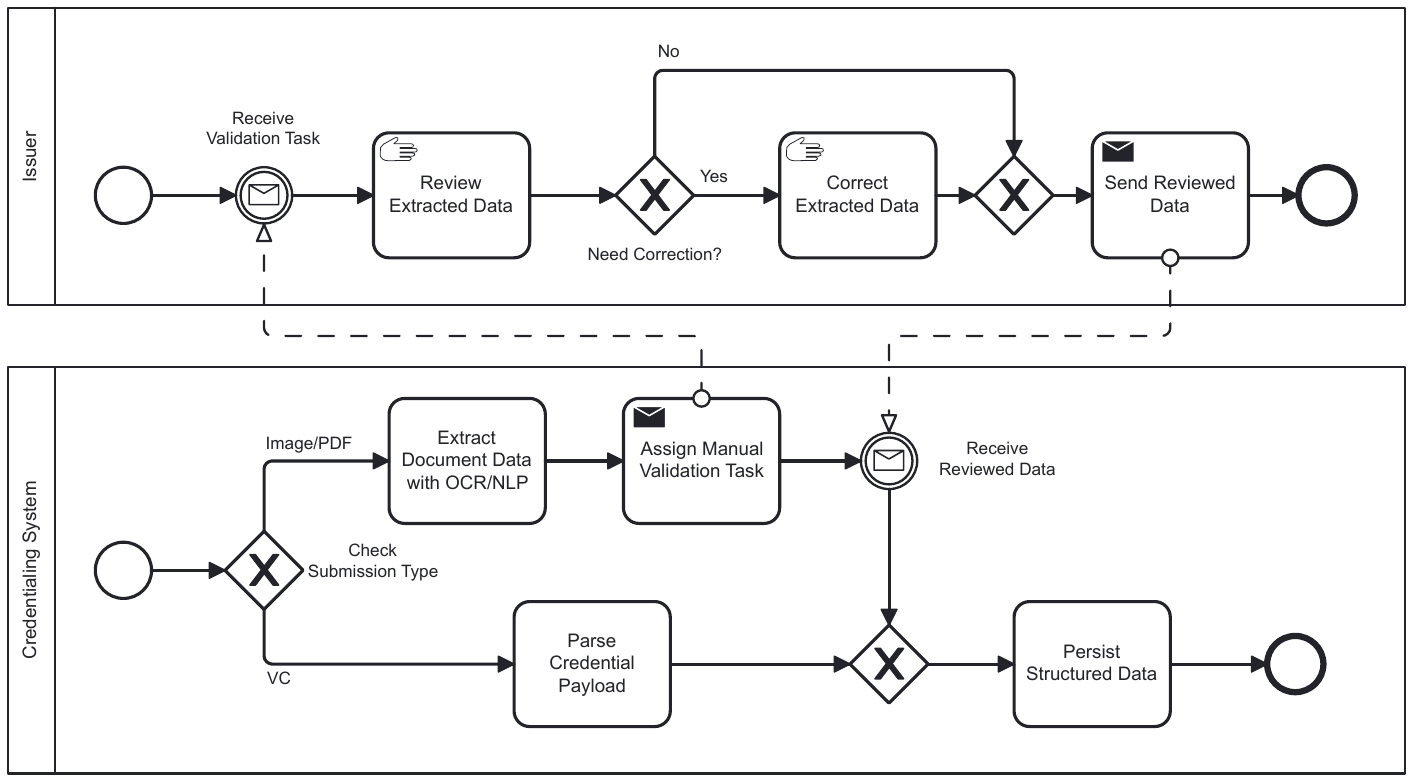}
  \caption{Stage 2 – Data Extraction and Correction Workflow}
  \label{fig:stage2_workflow}
\end{figure}

\paragraph{Stage 2 – Data Extraction and Correction}
For file-based submissions, the Document Processing Service applies OCR and NLP techniques to extract key fields such as names, addresses, or identifiers. The Issuer reviews this structured output and may correct errors or omissions. If the input was already a VC, its content is parsed directly without further validation (Figure~\ref{fig:stage2_workflow}).  

\begin{figure}[tb]
  \centering
  \includegraphics[width=\columnwidth]{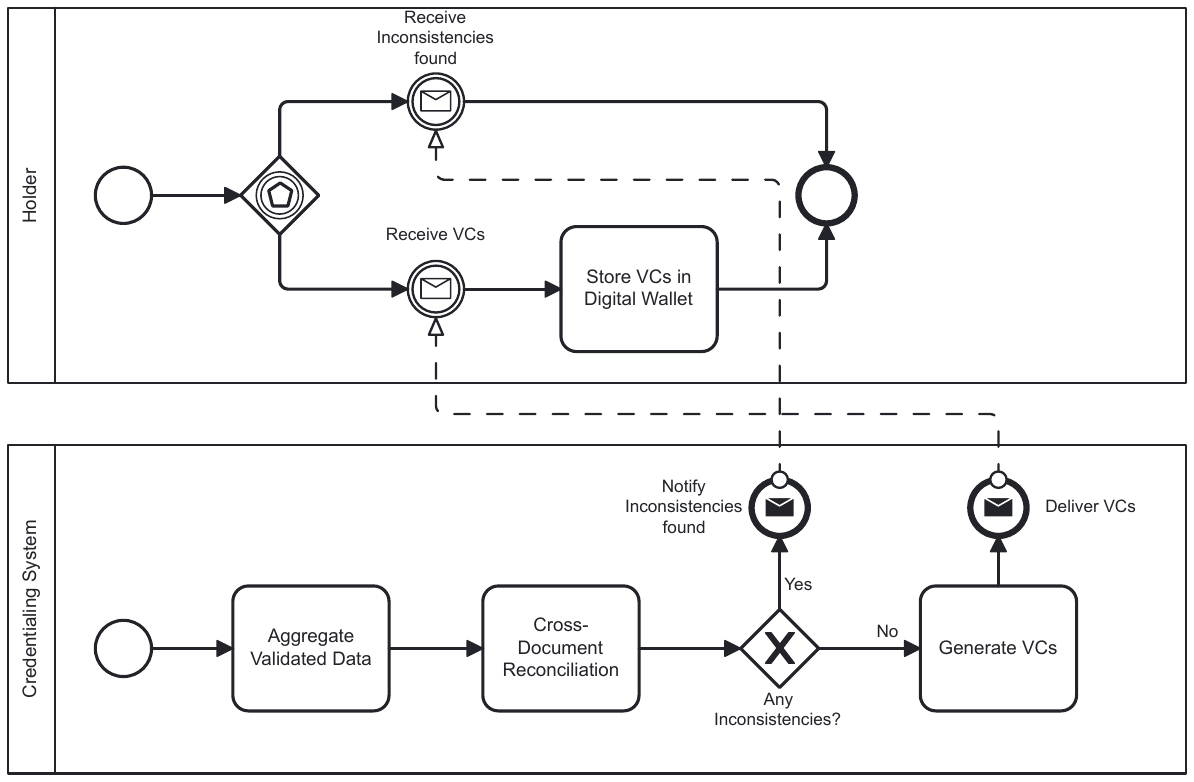}
  \caption{Stage 3 – Data Reconciliation and Credential Issuance Workflow}
  \label{fig:Stage3BPMN}
\end{figure}

\paragraph{Stage 3 – Data Reconciliation and Issuance}
The system checks the validated data across documents to detect inconsistencies (for example, mismatched names or missing references). If all data is consistent, the Issuer generates a VC and signs it cryptographically before delivering it to the Holder (Figure~\ref{fig:Stage3BPMN}).  

These three stages together define how heterogeneous inputs are progressively transformed into portable and verifiable digital credentials.  

\subsubsection{Post-Issuance Workflows}
Once credentials have been issued, they enter their active life cycle, where two complementary processes ensure that they remain reliable and trustworthy over time.  

\begin{figure}[tb]
  \centering
  \includegraphics[width=\columnwidth]{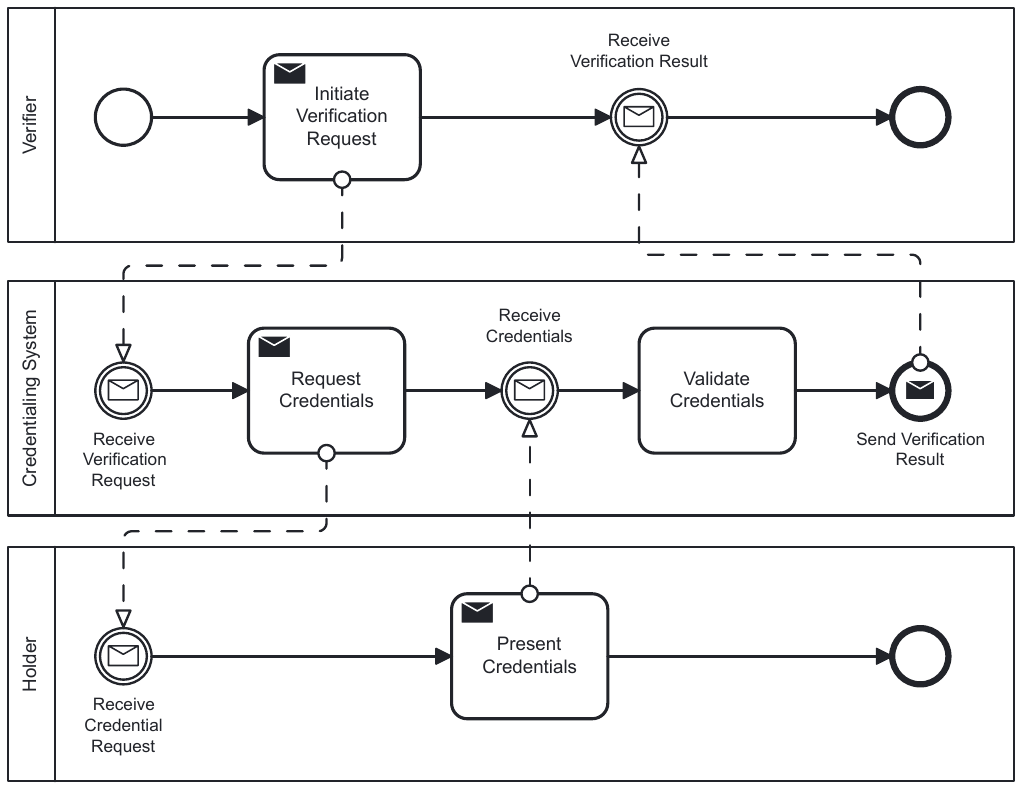}
  \caption{Credential Verification Workflow}
  \label{fig:credentialVerification}
\end{figure}

\paragraph{Verification}
When a Holder presents a credential, a Verifier must confirm its authenticity. This involves checking the Issuer’s digital signature, validating the integrity of the credential, and consulting the status registry to ensure that it has not been revoked. Verification can occur either through the Credentialing System, which offers a unified interface for multi-credential checks, or independently with any standards-compliant verifier. Figure~\ref{fig:credentialVerification} illustrates this process, showing how Holders, Verifiers, and the system interact to confirm validity.  

\begin{figure}[tb]
  \centering
  \includegraphics[width=\columnwidth]{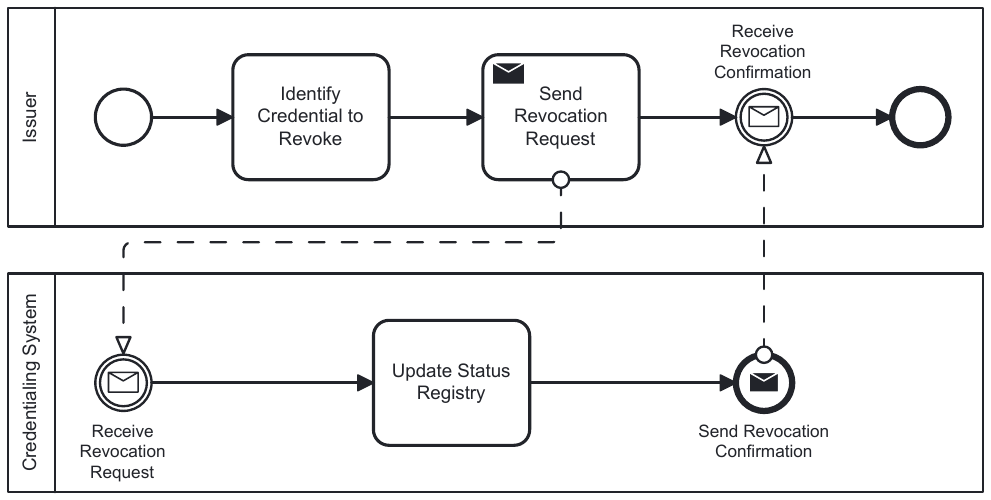}
  \caption{Credential Revocation Workflow}
  \label{fig:credentialRevocation}
\end{figure}

\paragraph{Revocation} 
Credentials may need to be invalidated when underlying data changes (e.g., contract expiration), when fraudulent activity is detected, or when errors are discovered. The Issuer initiates revocation by updating the credential’s status in the registry, ensuring that future verification requests reflect its invalid state. This process maintains trust by preventing the reuse of outdated or compromised credentials. Figure~\ref{fig:credentialRevocation} shows the main steps of revocation, from Issuer request to status update and confirmation.  

Together, verification and revocation close the credential life cycle. They ensure that once issued, credentials can be relied upon in dynamic real-world processes, while still allowing their status to be updated if conditions change.

\subsection{Abstract Architecture}
\begin{figure}[tb]
    \centering
    \includegraphics[width=\columnwidth]{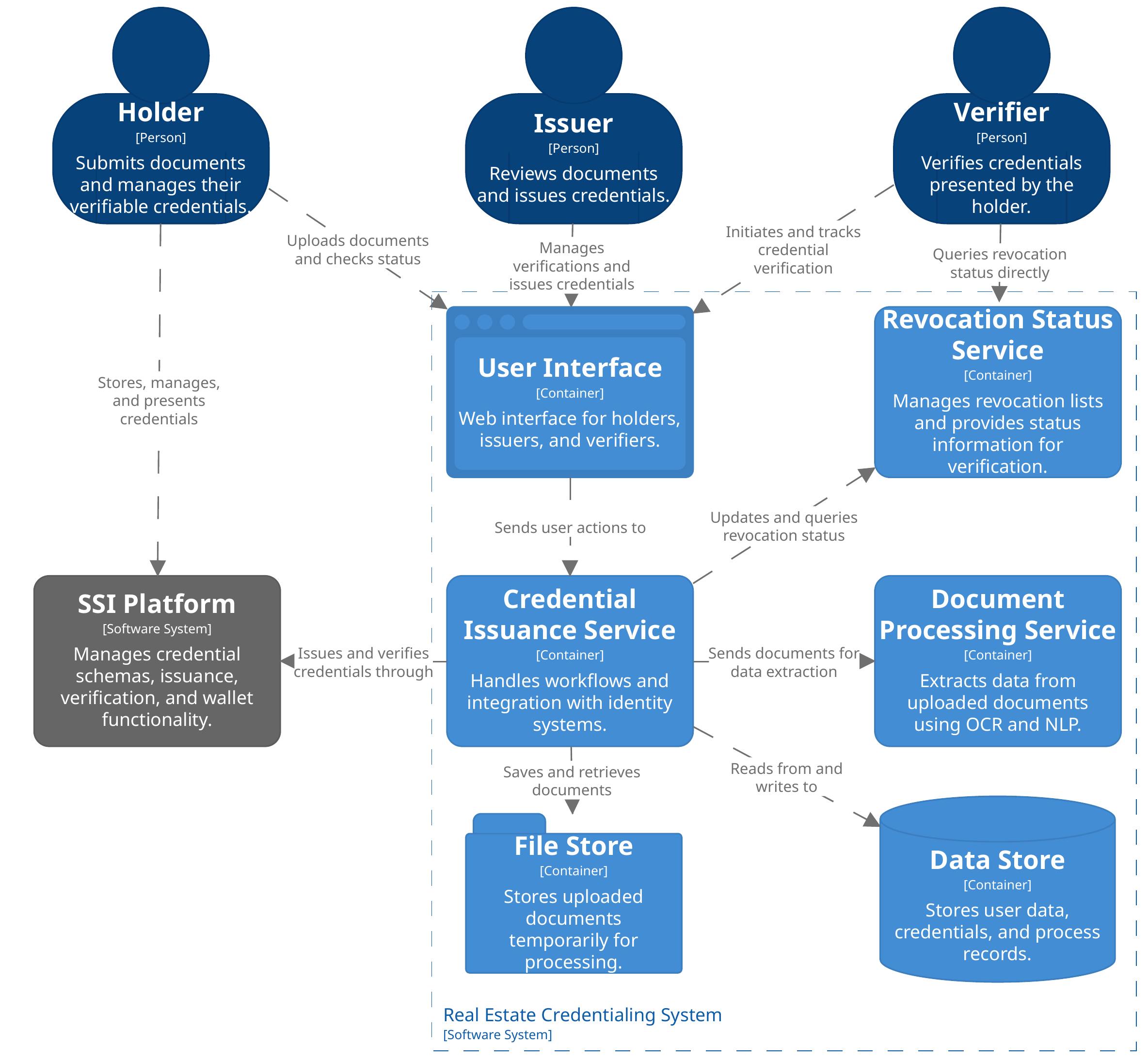}
    \caption{System Design – C4 Container Diagram}
    \label{fig:system-architecture}
\end{figure}

The architecture in Figure~\ref{fig:system-architecture} aligns actors and workflows with modular services.  
A web interface supports all roles, while a \emph{Document Processing Service} handles OCR and NLP, a \emph{Credential Issuance Service} orchestrates business logic, and a \emph{Revocation Status Service} maintains validity.  
Data and file stores persist inputs and outputs, and an external SSI platform provides DID resolution, signing, and wallet interoperability.  
This modular design ensures extensibility and interoperability while preserving decentralized verification.

\subsection{Design Considerations}
The design of the \textbf{Credentialing System} was guided by several technical and architectural decisions that ensure reliability, interoperability, and scalability. Four aspects were particularly critical:

\paragraph{Document Data Extraction Pipeline}
Reliable structured data extraction is essential, but annotated real-estate datasets are scarce. To address this, the system employs a synthetic dataset strategy in which document templates are populated with realistic but fictitious values, automatically annotated, and augmented with noise to mimic real-world imperfections. A fine-tuned NLP model processes the OCR output into key–value pairs, supporting accuracy while avoiding privacy concerns.

\paragraph{Decentralized Identifiers}
All actors (Holders, Issuers, Verifiers) are represented by blockchain-anchored DIDs. This choice provides tamper-evident persistence, decentralized key management, and compatibility with standard verifier tools. Lightweight methods such as \texttt{did:key} were excluded due to limited persistence and lack of service endpoints.

\paragraph{Revocation Strategy}
Because real estate documents often expire or change, management of the life cycle of credentials is essential. The system adopts a status list approach, where a dedicated service issues verifiable status credentials. This allows for scalable and privacy-preserving verification, supports multiple states (valid, revoked, suspended), and works in both online and offline scenarios.

\paragraph{Credential Formats and SSI Platform}
To ensure interoperability, credentials are represented in JSON-LD with Linked Data Signatures, aligning with W3C recommendations and supporting semantic extensibility. The architecture remains open to more advanced privacy-preserving formats such as BBS+ in the future. For infrastructure, walt.id was selected as the SSI platform, given its strong standards compliance, DID method support, and modular open-source tooling, although alternatives could be integrated if needed.

\section{System Implementation}
\label{sec:implement}
This section describes how the \textbf{Credentialing System} was implemented as a functional prototype. The implementation is organized into four areas: the overall architecture, the data extraction pipeline, the credential issuance service, and the user interface. Together, these components enable the end-to-end life cycle of real-estate credentials, from document ingestion to decentralized verification.

\subsection{Architecture}
\begin{figure}[tb]
    \centering
    \includegraphics[width=\columnwidth]{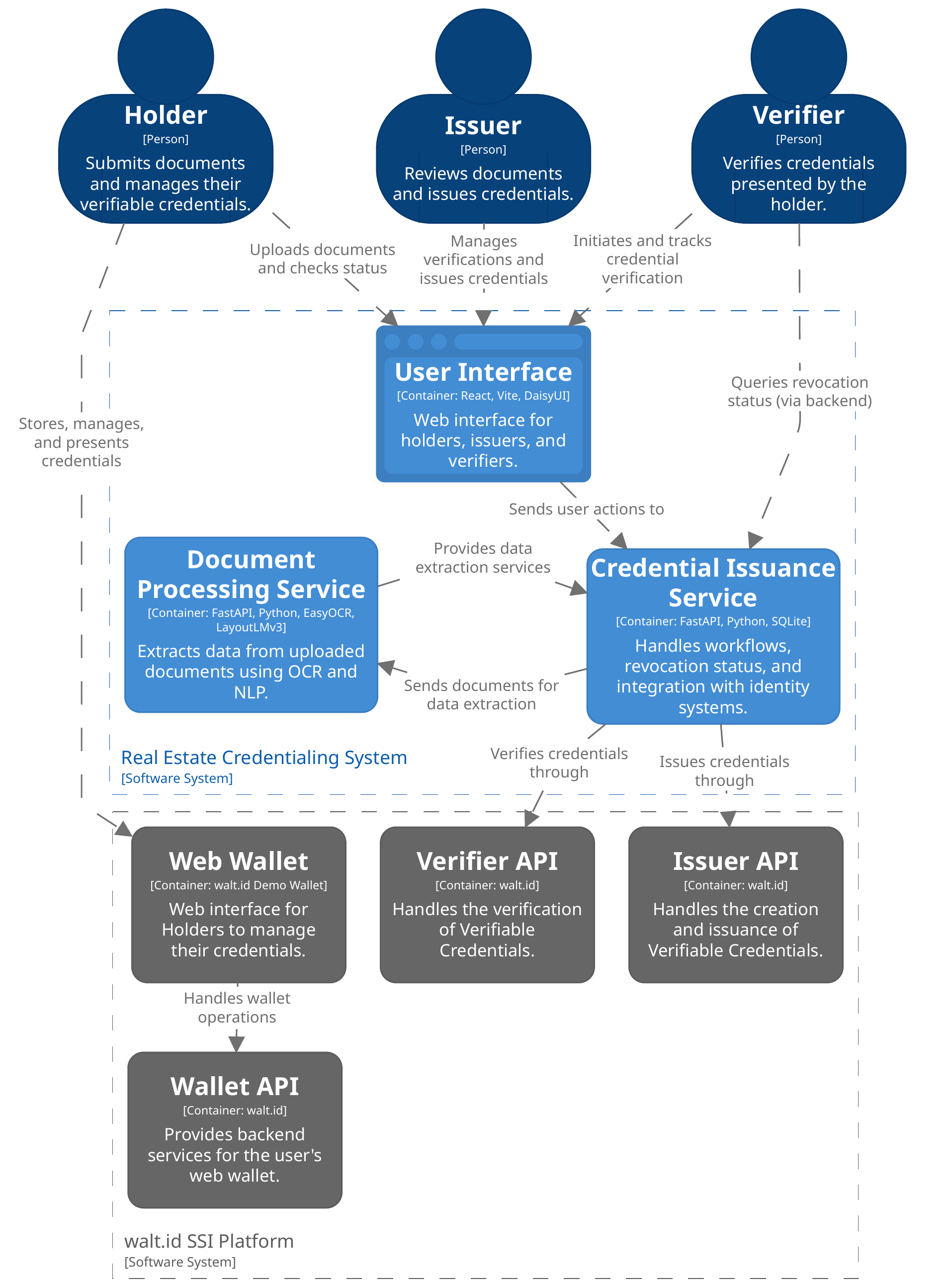}
    \caption{System architecture (C4 container diagram with technology stack)}
    \label{fig:impl-c4-container}
\end{figure}

The prototype follows a modular architecture where independent services are deployed and integrated with an external SSI provider. Figure~\ref{fig:impl-c4-container} illustrates the container-level design.

The containers and their responsibilities are:
\begin{itemize}[leftmargin=\parindent]
    \item \textit{User Interface (React + Vite + Tailwind/DaisyUI):} provides role-based interaction for Holders, Issuers, and Verifiers, including document submission, validation, issuance, and verification.
    \item \textit{Document Processing Service (FastAPI + EasyOCR + LayoutLMv3):} implements OCR and NLP to transform heterogeneous inputs into structured fields.
    \item \textit{Credential Issuance Service (FastAPI + SQLite):} coordinates ingestion, validation, reconciliation, issuance, and revocation. In this prototype, local storage is embedded for simplicity.
    \item \textit{SSI Platform (walt.id):} provides a decentralized identity infrastructure, including APIs for credential issuance (OID4VCI) and presentation (OID4VP).
\end{itemize}

This separation ensures scalability: each component can be replaced or extended independently, while interoperability with external SSI platforms guarantees that the prototype aligns with open standards.

\subsection{Data Extraction Pipeline}
A central capability of the system is to convert heterogeneous real-estate documents into structured fields. The pipeline consists of OCR transcription, entity recognition, and post-processing. To achieve reliable performance despite scarce annotated data, the implementation combines off-the-shelf tools with synthetic dataset generation and transformer fine-tuning.

\subsubsection{OCR and NLP Tools}
EasyOCR was chosen for OCR because of its simplicity and acceptable performance on a set of documents currently used in Portugal. It produces tokens, bounding boxes, and confidences that serve as input to NLP.  
For entity recognition, LayoutLMv3 was selected as it jointly models text, layout, and visual features, making it effective for structured documents such as IDs and property records.

\subsubsection{Synthetic Dataset Generation}
\begin{figure}[tb]
    \centering
    \subfigure[Blank Citizen Card Template]{
        \label{fig:template_blank}
        \includegraphics[width=0.45\columnwidth]{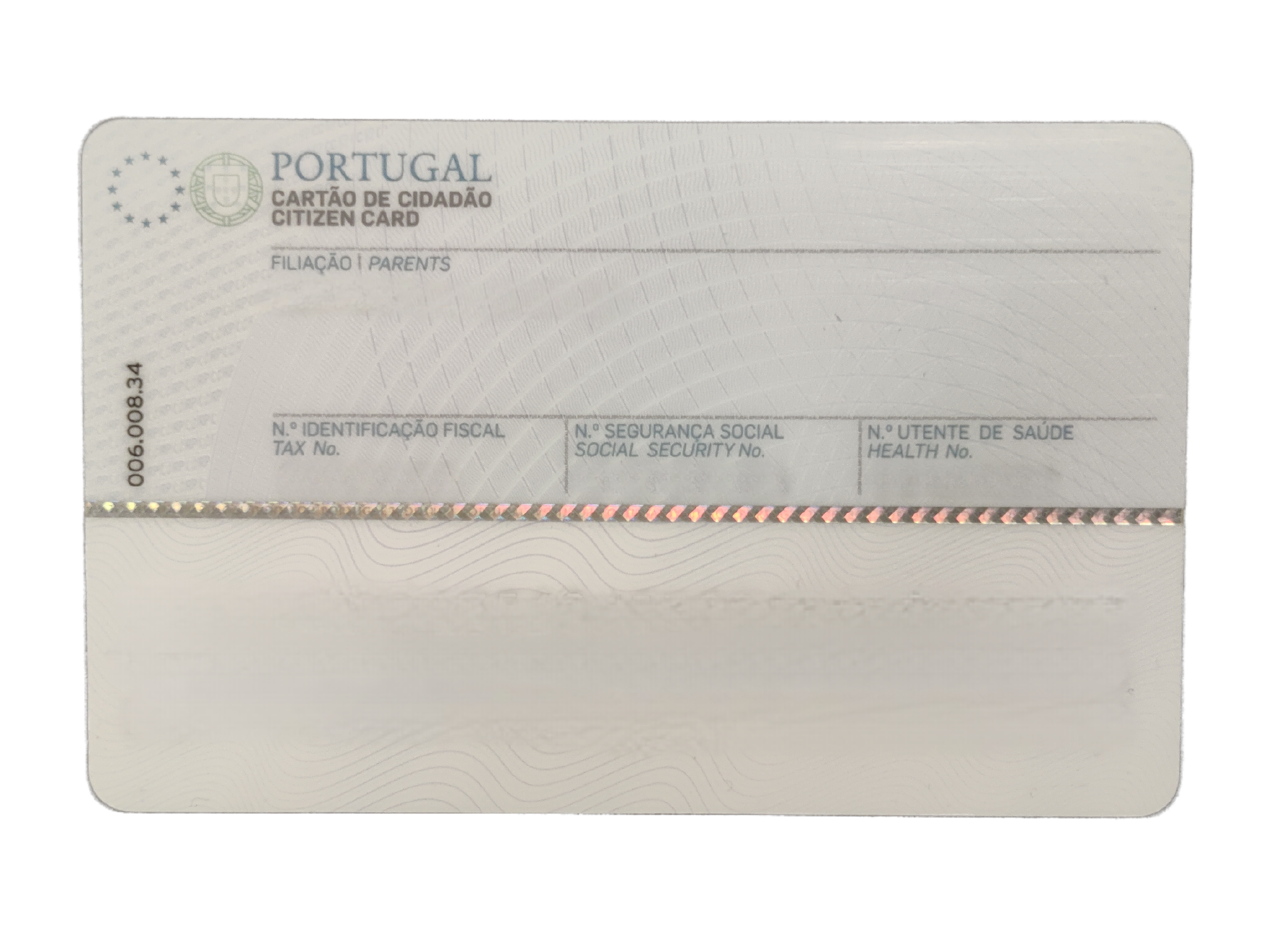}
    }
    \subfigure[Template Filled with Fake Data]{
        \label{fig:template_filled}
        \includegraphics[width=0.45\columnwidth]{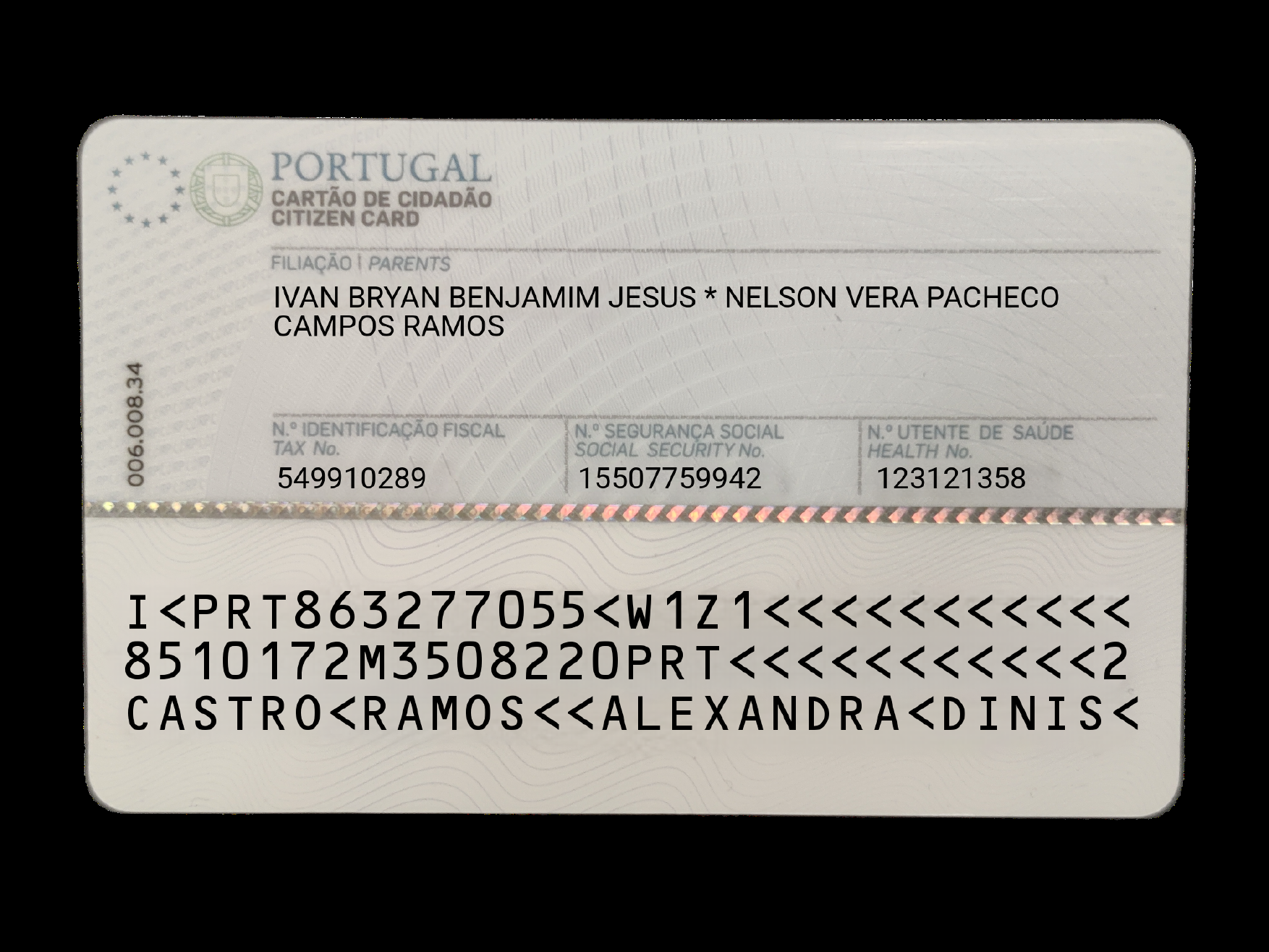}
    }
    \subfigure[Augmented Image with Noise and Distortion]{
        \label{fig:template_augmented}
        \includegraphics[width=0.45\columnwidth]{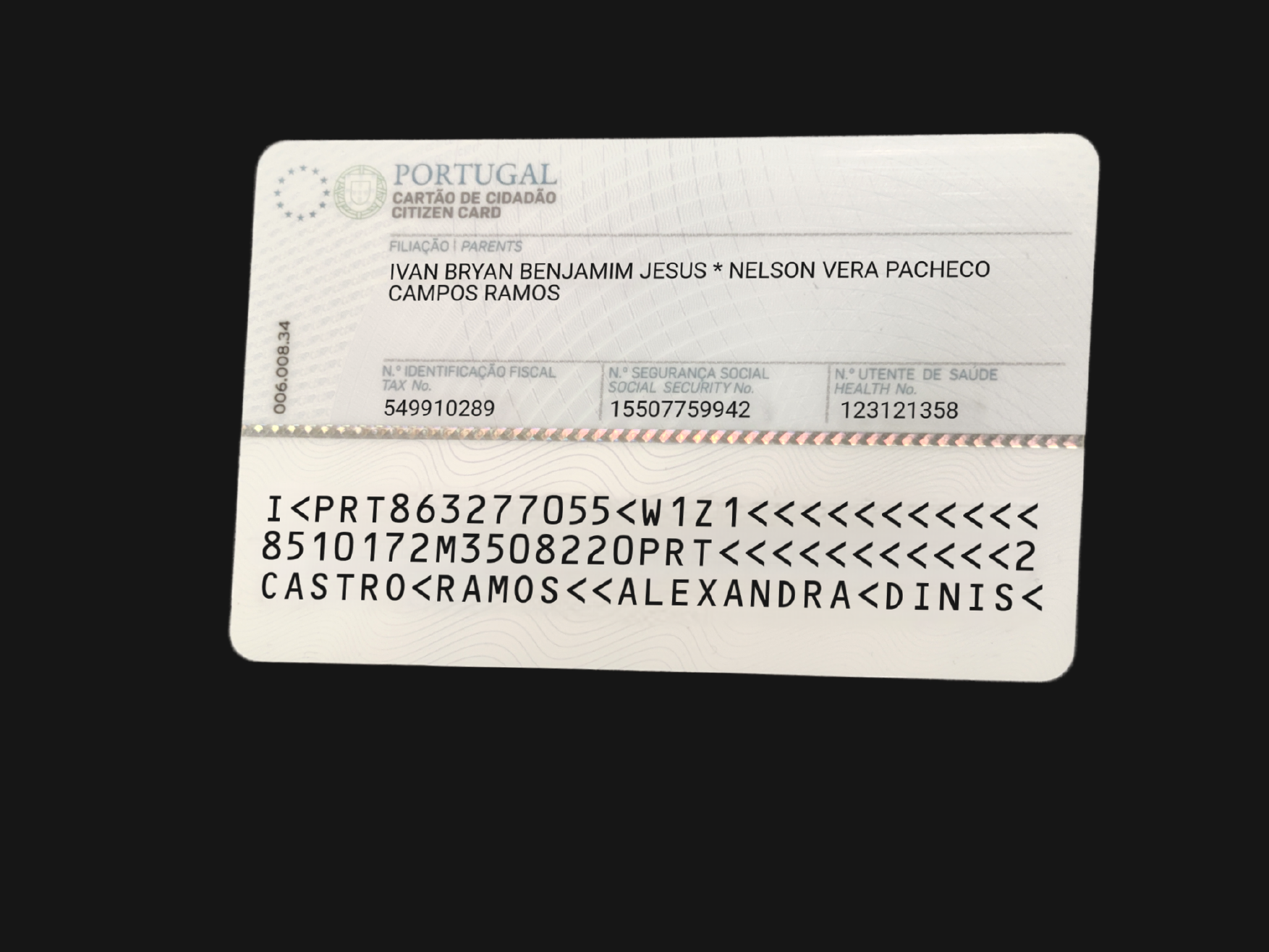}
    }
    \subfigure[Final Synthetic Sample Ready for OCR]{
        \label{fig:template_final}
        \includegraphics[width=0.45\columnwidth]{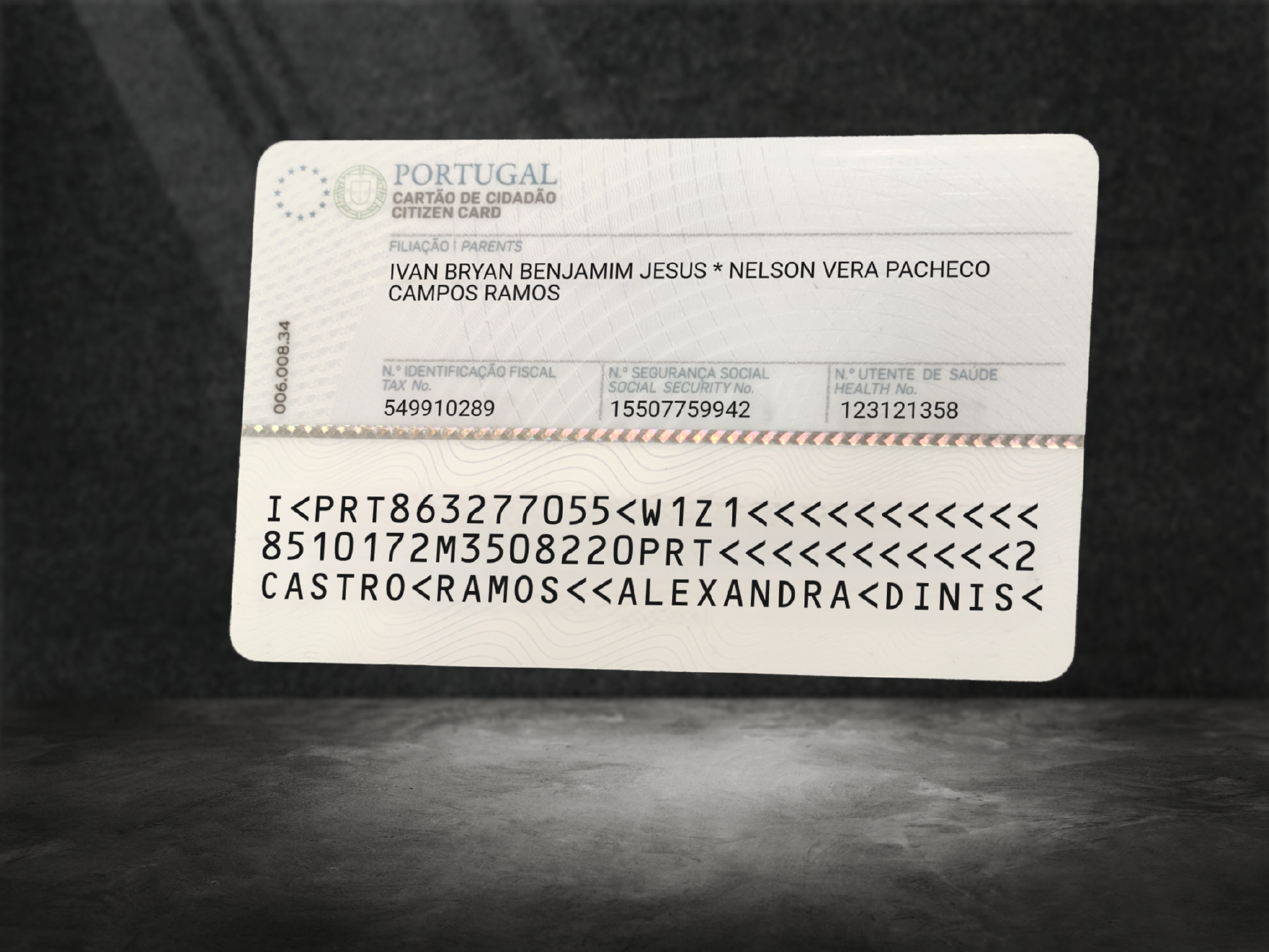}
    }
    \caption{Key steps in generating a synthetic Citizen Card sample.}
    \label{fig:synthetic_pipeline}
\end{figure}

Because no large labeled corpora of Portuguese real estate documents are publicly available, three synthetic datasets were created that cover Citizen Cards, Energy Certificates, and Property Records. The generation process followed four main steps. First, document templates were manually designed to reproduce the structure and layout of real forms. Second, these templates were populated with realistic but fictitious values (e.g., names, addresses, identifiers) generated using \emph{Faker} and rendered with \emph{Pillow}. Third, visual augmentations such as blur, rotation, and noise were applied with \emph{Albumentations} to simulate scanning artifacts and inconsistencies. Finally, each sample was automatically annotated: bounding boxes and field labels were recorded during rendering, then aligned with OCR outputs using intersection-over-union (IoU). Figure~\ref{fig:synthetic_pipeline} illustrates the main stages of this pipeline.

\begin{figure}[th]
\centering
  \includegraphics[width=0.7\columnwidth]{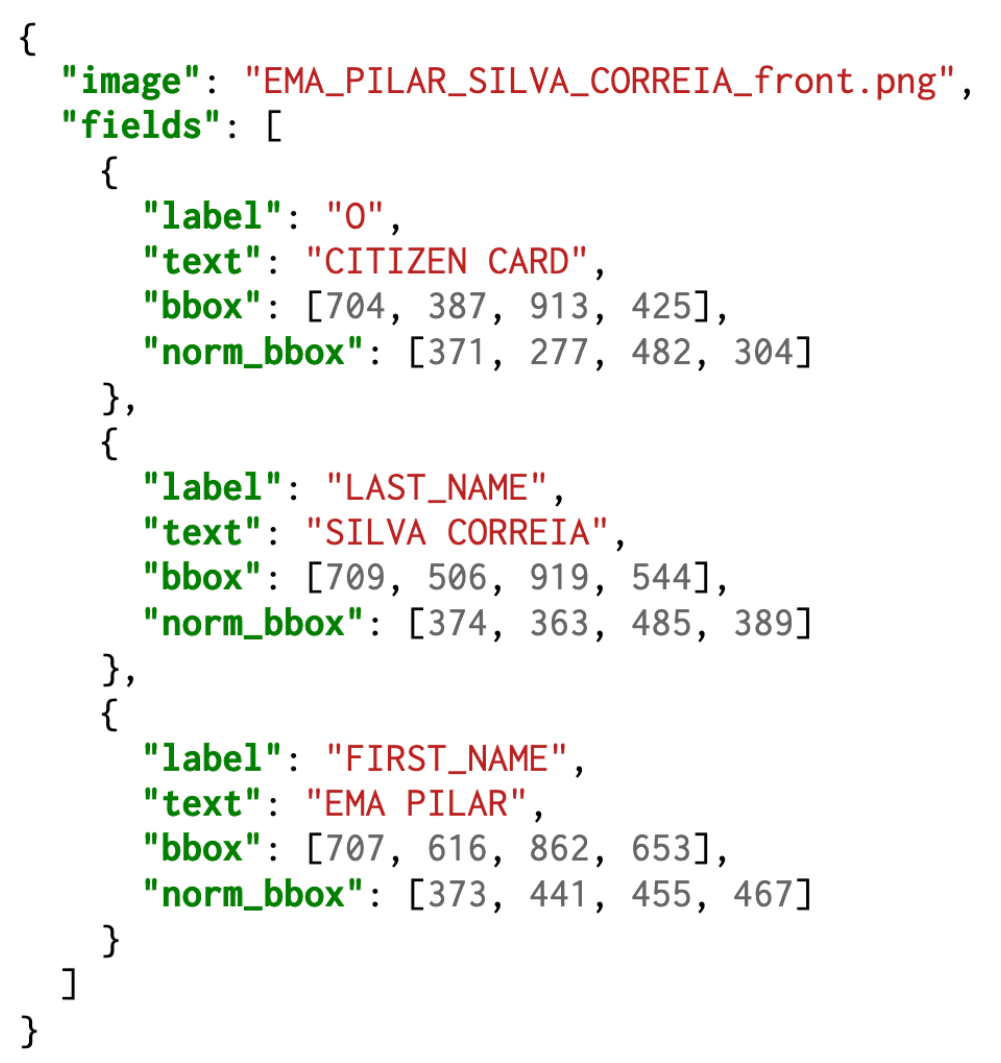}
\caption{Sample annotation JSON file}
\label{lst:sample_json}
\end{figure}

The output of this process is not only a rendered document image, but also a structured annotation file that links text, labels, and bounding boxes. A sample JSON annotation is shown in Figure~\ref{lst:sample_json}, illustrating how each entity (e.g., \texttt{FIRST\_NAME}, \texttt{LAST\_NAME}) is paired with its text content and coordinates. These annotations serve as ground truth for model training and evaluation, while maintaining privacy by avoiding the use of real data.

\subsubsection{Model Fine-Tuning}
LayoutLMv3 was fine-tuned separately for each document type. Training used the Hugging Face \texttt{Transformers} framework with BIO tagging, sliding windows for long sequences, and entity-level evaluation metrics. Despite modest hardware (RTX 2070 GPU), convergence was achieved in five epochs, demonstrating feasibility in constrained environments.  
Table~\ref{tab:layoutlm_hyperparams} summarizes the main hyperparameters.

\begin{table}[htbp]
\centering
\rowcolors{2}{lightgray}{white}
\begin{tabular}{lll}
\toprule
\textbf{Parameter} & \textbf{Value} & \textbf{Rationale} \\
\midrule
Learning rate   & 2e-5  & Stable fine-tuning on small datasets. \\
Batch size      & 2     & Fits GPU memory, ensures stability. \\
Epochs          & 5     & Converges without overfitting. \\
Weight decay    & 0.01  & Regularization. \\
Seq. length     & 512   & Covers typical documents. \\
Precision       & FP16  & Efficient GPU usage. \\
\bottomrule
\end{tabular}
\caption{LayoutLMv3 fine-tuning setup}
\label{tab:layoutlm_hyperparams}
\end{table}

\subsubsection{Integration}
The extraction pipeline is deployed as a FastAPI microservice, exposing a simple interface for document processing. Submissions can be images or PDFs, which are internally normalized and passed through the OCR engine before being processed by the fine-tuned LayoutLMv3 model. The output is returned as structured JSON, with fields such as \texttt{sex} or \texttt{registry\_number} mapped directly to their extracted values. The format supports arrays to represent repeated entities, for example multiple property owners or co-holders.

Several design heuristics improve robustness in real-world use. Token order inconsistencies from OCR are mitigated with custom sorting rules, while sliding-window partitioning ensures that long or complex layouts remain processable without exceeding model limits. Noise resilience is enhanced by training on synthetically augmented samples, simulating distortions, font variations, and common OCR misrecognitions. Together, these strategies allow the pipeline to generalize beyond idealized inputs and maintain reliable accuracy in practice.

\subsection{Credential Issuance Service}
\begin{figure}[tb]
    \centering
    \includegraphics[width=\columnwidth]{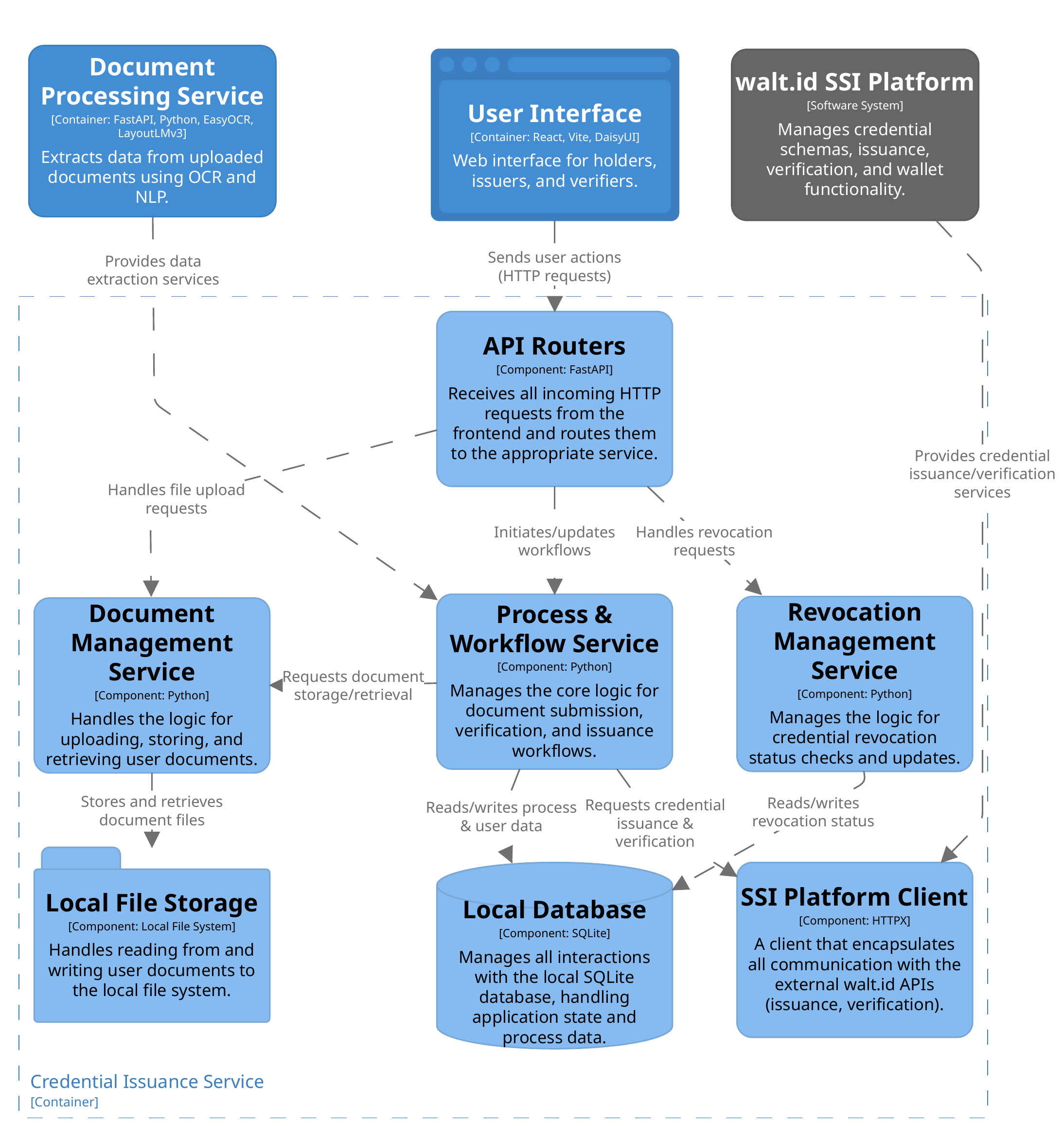}
    \caption{Component architecture of the Credential Issuance Service}
    \label{fig:c4_backend}
\end{figure}

The credential issuance service connects the extraction pipeline with credential life cycle management. It handles ingestion, validation, reconciliation, issuance, and revocation. Figure~\ref{fig:c4_backend} shows the component-level architecture.

\paragraph{Architecture.} Implemented in FastAPI, the service exposes endpoints via \textit{API Routers}. A \textit{Process \& Workflow Service} coordinates the process stages, the \textit{Document Management Service} stores uploads, and \textit{SQLite Database} provides persistence. The \textit{Revocation Management Service} maintains status list entries. Integration with walt.id enables credential issuance and verification. While simplified for the prototype, the design allows individual modules (e.g., storage, revocation) to be externalized in production.

\paragraph{life cycle flows.}
\begin{itemize}[leftmargin=\parindent]
    \item \textit{Ingestion:} Holders upload documents or share existing credentials. The extraction is triggered asynchronously.
    \item \textit{Validation:} Issuers review and correct extracted data, or reject faulty submissions.
    \item \textit{Reconciliation:} Approved data is cross-checked across documents, such as NIF consistency, name matching, or verifying Energy Certificate coordinates against Property Records.
    \item \textit{Issuance:} If validation and reconciliation succeed, the service calls the walt.id issuer API to create a credential offer URL, which the Holder redeems in their wallet.
    \item \textit{Verification:} Credentials can later be presented to a verifier, who validates digital signatures, schema compliance, and validity periods. These checks may be performed locally or delegated to a verifier API. However, the revocation status is always determined from the \texttt{credentialStatus} field of the credential, which points to a VC status list that the verifier retrieves and inspects.
    \item \textit{Revocation:} Issuers may revoke individual credentials or all credentials tied to a process by updating the relevant status bits. Because revocation information is published in a signed status list credential, verifiers can confirm validity in a decentralized way without continuous communication with the issuer.
\end{itemize}

This workflow blends automation with human oversight, ensuring both accuracy and trustworthiness while adhering to interoperable SSI standards.

\subsection{User Interface}
The user interface provides the interaction layer of the prototype, enabling Holders, Issuers, and Verifiers to participate in credential workflows. Although deliberately lightweight, it demonstrates the entire life cycle in an accessible way.

\paragraph{Architecture and Technologies.}
The frontend was implemented using React with Vite for fast builds, styled with Tailwind CSS and DaisyUI. This stack was chosen to balance simplicity and clarity with modern web development practices. A dedicated API module centralizes communication with the backend, while reusable utilities handle tasks such as process state translation or bitstring decoding for revocation checks. Although minimal, the architecture supports modular extensions and a clear separation of concerns.

\paragraph{Role-based Views.}
The interface separates the three main user roles into dedicated views:
\begin{itemize}[leftmargin=\parindent]
    \item \textit{Holder:} Starts new credentialing processes, uploads documents or shares existing credentials, and redeems credential offers in their wallet. The dashboard presents active processes and their current status, ensuring transparency over progress.
    \item \textit{Issuer:} Reviews extracted data, validates or corrects submissions, and decides whether to approve or reject them. This human-in-the-loop step provides a safeguard against OCR or model errors, ensuring accuracy before issuance.
    \item \textit{Verifier:} Requests credential presentations and validates them. Verification runs partly in the browser, showing that cryptographic checks and revocation validation can occur without backend dependencies, reinforcing decentralization.
\end{itemize}

\begin{figure}[tb]
  \centering
  \subfigure[Holder dashboard]{
    \includegraphics[width=0.8\columnwidth]{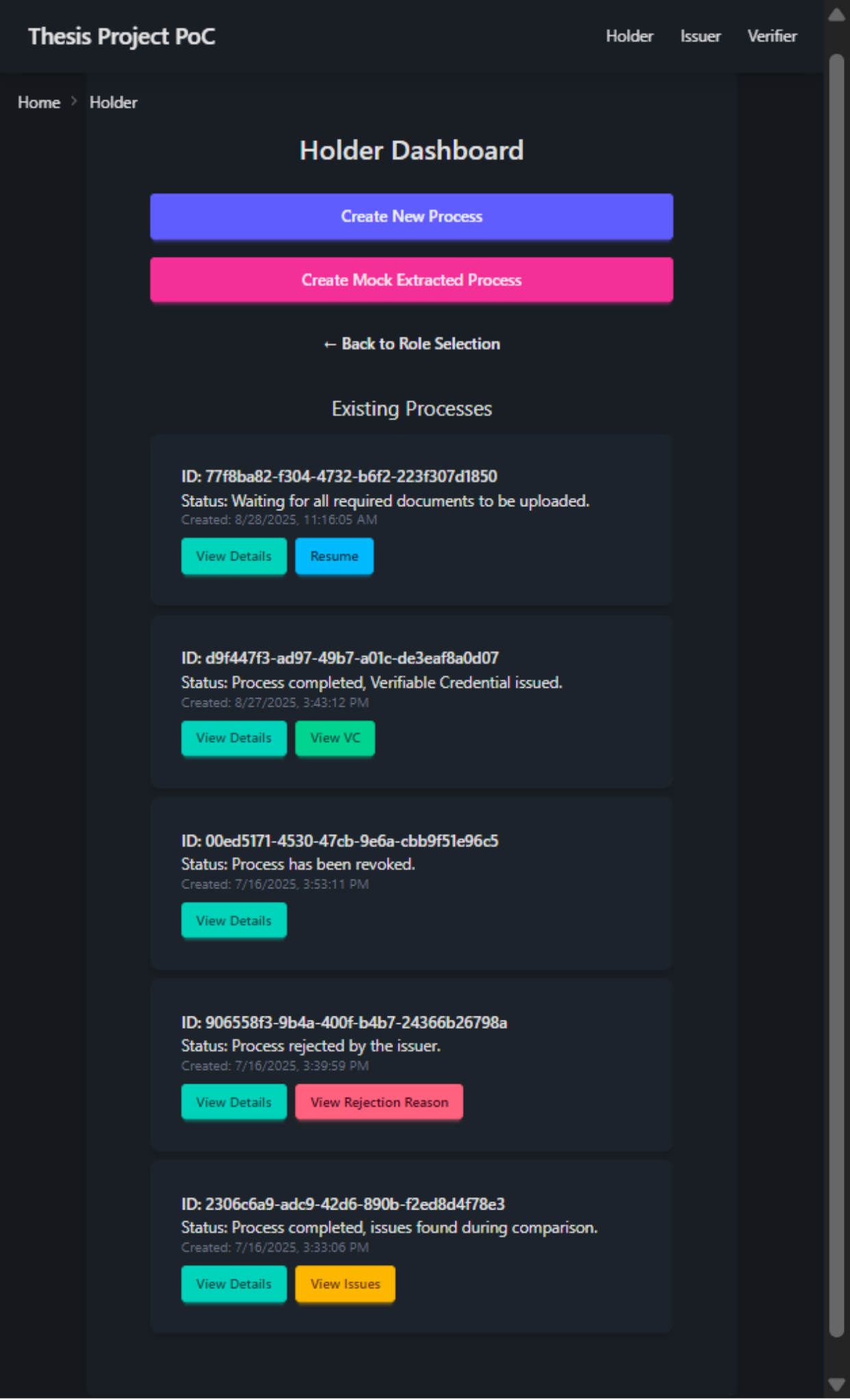}
    \label{fig:ui_holder}
  }
  \subfigure[Issuer validation view]{
    \includegraphics[width=0.45\columnwidth]{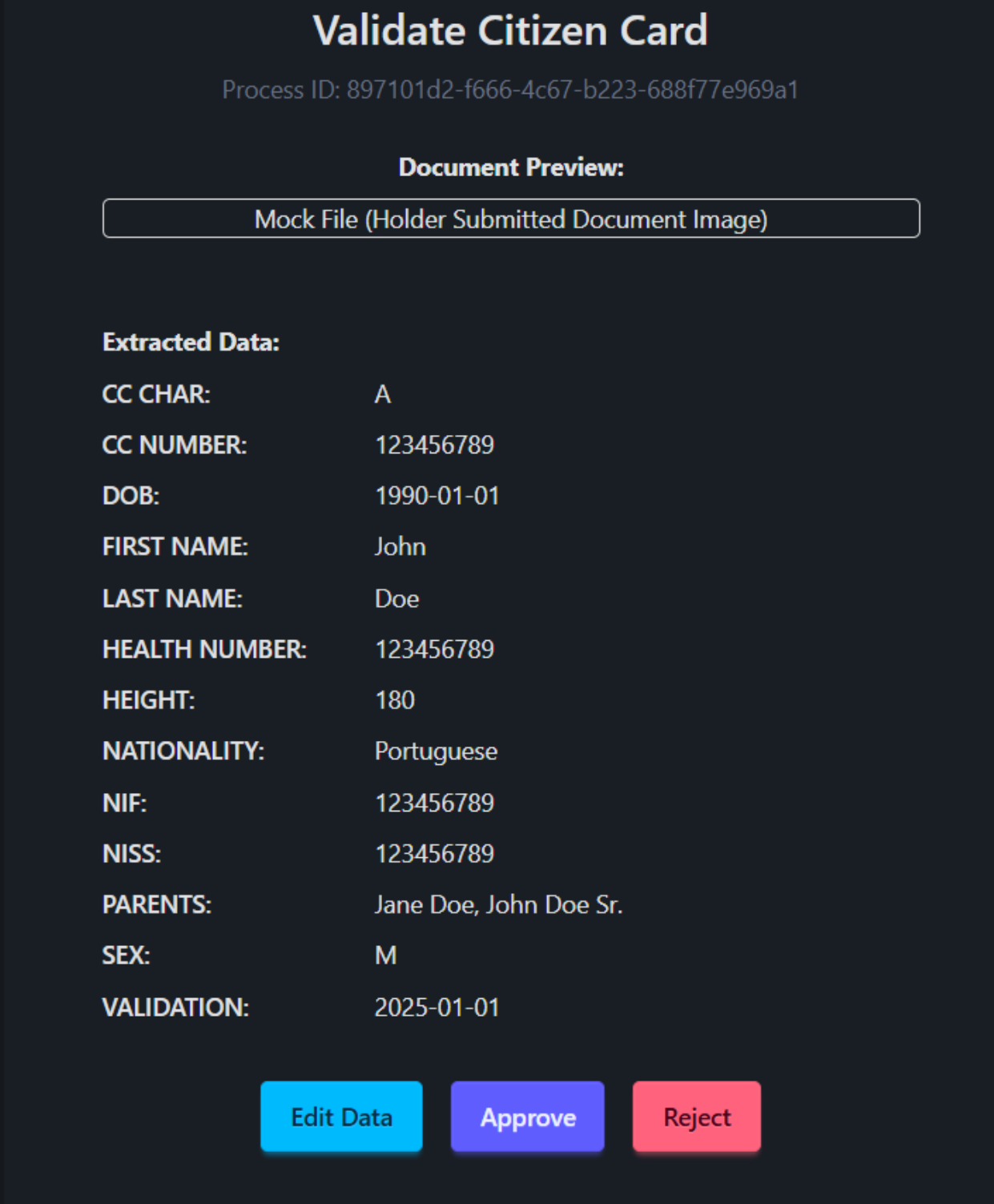}
    \label{fig:ui_issuer}
  }
  \subfigure[Verifier dashboard]{
    \includegraphics[width=0.45\columnwidth]{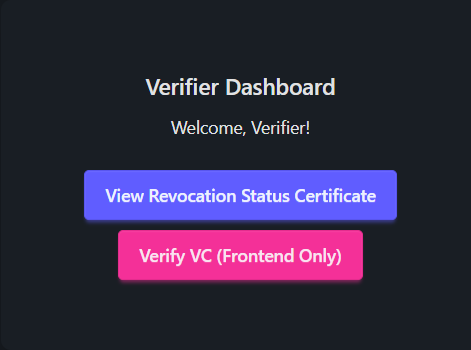}
    \label{fig:ui_verifier}
  }
  \caption{User interface views for the three roles: Holder, Issuer, and Verifier.}
  \label{fig:user_interface}
\end{figure}

Examples of the three roles are shown in Figure~\ref{fig:user_interface}.

\paragraph{Design Considerations.}
The interface emphasizes clarity over completeness. The screens are intentionally simplified to highlight the critical actions of each role rather than replicate a production-grade platform. This choice reduces complexity while keeping the focus on demonstrating end-to-end interoperability with the backend and SSI platform. At the same time, the modular frontend design makes it possible to evolve into a richer application by adding workflows such as multi-issuer coordination or extended revocation management.

\paragraph{Demonstration Value.}
By simulating all major actors and flows, the interface bridges technical backend services with end-user interaction. It highlights how documents are transformed into VCs, how issuers exercise oversight, and how verifiers can independently validate trust. The frontend therefore serves as a key tool for demonstrating the prototype in academic and industry settings.

\section{Evaluation}
\label{sec:evaluation}

This section evaluates the proposed system to determine its effectiveness, limitations, and readiness for deployment in real-world document processing scenarios. It outlines the experimental setup, describes the evaluation metrics, and presents results from model testing, end-to-end evaluation, and comparison with human annotators.

\subsection{Experimental Setup}
All experiments were conducted in a controlled local environment to ensure consistency and reproducibility. The hardware used included an AMD Ryzen 7 2700X processor, an NVIDIA RTX 2070 GPU (8 GB VRAM), 16 GB of RAM, and a 250 GB NVMe SSD, running Windows 11. 

The software environment was based on Python 3.10, managed with \verb|uv| for reproducible dependencies. Key libraries included \\ \verb|transformers|, \verb|torch|, \verb|easyocr|, and \verb|albumentations|.  

Datasets and trained model checkpoints are publicly available on the Hugging Face Hub\footnote{\url{https://huggingface.co/HenriqueLin}}, ensuring that the evaluation process can be fully replicated. Three synthetic datasets (Citizen Card, Energy Certificate, and Property Record) and their corresponding fine-tuned LayoutLMv3 models were used throughout the experiments.

\subsection{Evaluation Metrics}
The system was evaluated using Accuracy, Precision, Recall, and F1-score. For \textbf{NLP model evaluation}, entity-level metrics were computed with the \texttt{seqeval} library, comparing predicted and ground-truth spans. For the \textbf{end-to-end pipeline}, evaluation was performed at the field level, including OCR outputs, under three matching modes: exact (strict equality), tolerant (normalization + edit distance), and super tolerant (additional domain-specific normalization).

\subsection{Model Training and Testing}
\begin{figure}[tb]
    \centering
    \includegraphics[width=\columnwidth]{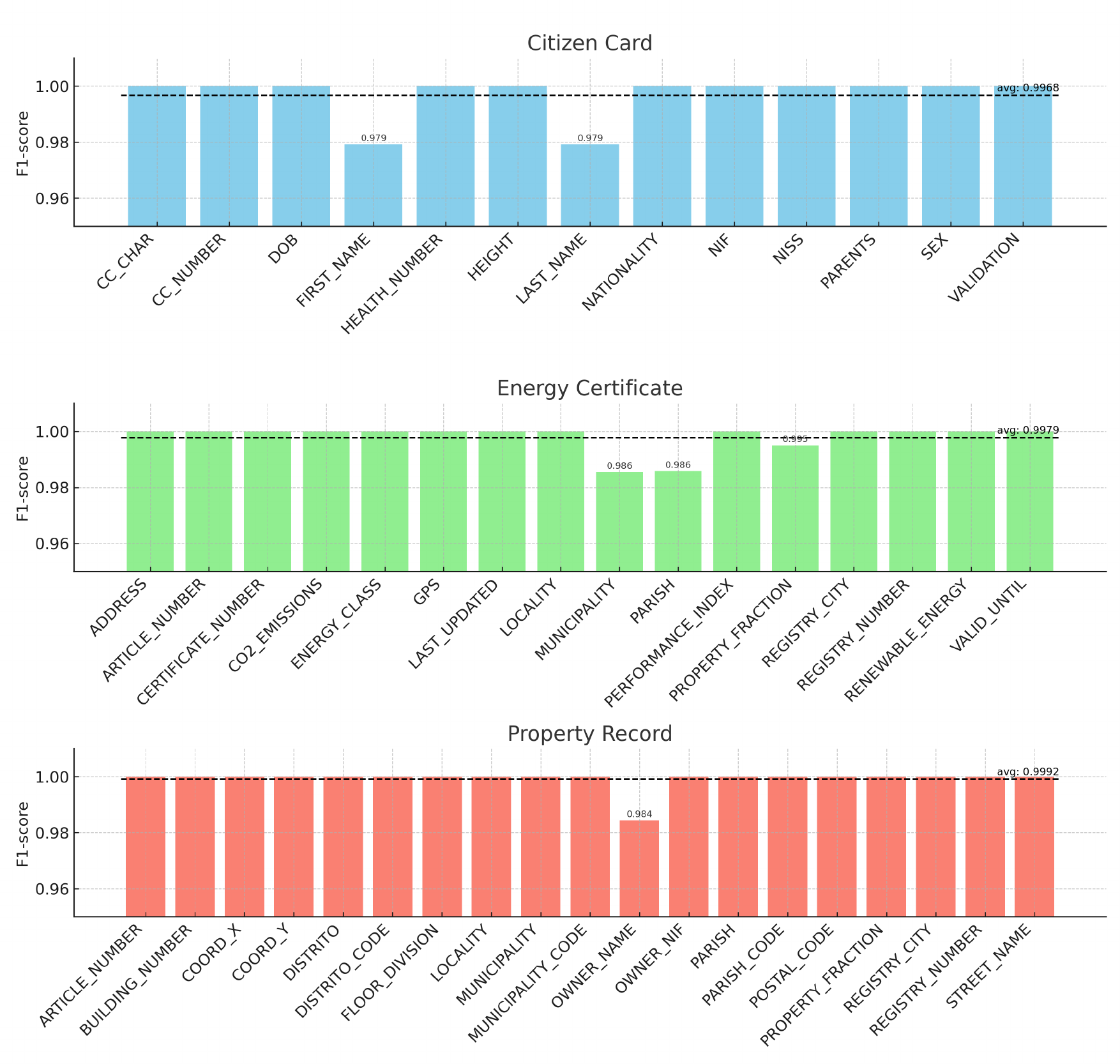}
    \caption{Per-field F1-scores on held-out test sets across document types}
    \label{fig:f1-field-level}
\end{figure}

Three LayoutLMv3 models were trained, one per document type. Training converged rapidly, with models reaching near-perfect results in five epochs. Testing on held-out synthetic sets confirmed this: average F1-scores reached 0.9968 (Citizen Card), 0.9979 (Energy Certificate), and 0.9992 (Property Record). As shown in Figure~\ref{fig:f1-field-level}, most fields scored above 0.98, with only slight variation in names and address-related fields.

These results confirm that synthetic datasets provide a strong basis for training document-specific entity extraction models.

\subsection{End-to-End Pipeline Performance}
To assess practical performance, the complete OCR+NLP pipeline was tested on 50 synthetic samples per document type. Results were measured at the field level, with increasing tolerance improving F1-scores across all types. Citizen Card achieved the highest performance (up to 0.9628 F1), while Energy Certificates and Property Records had slightly lower results due to higher layout complexity and OCR noise.

\begin{figure}[tb]
    \centering
    \includegraphics[width=\columnwidth]{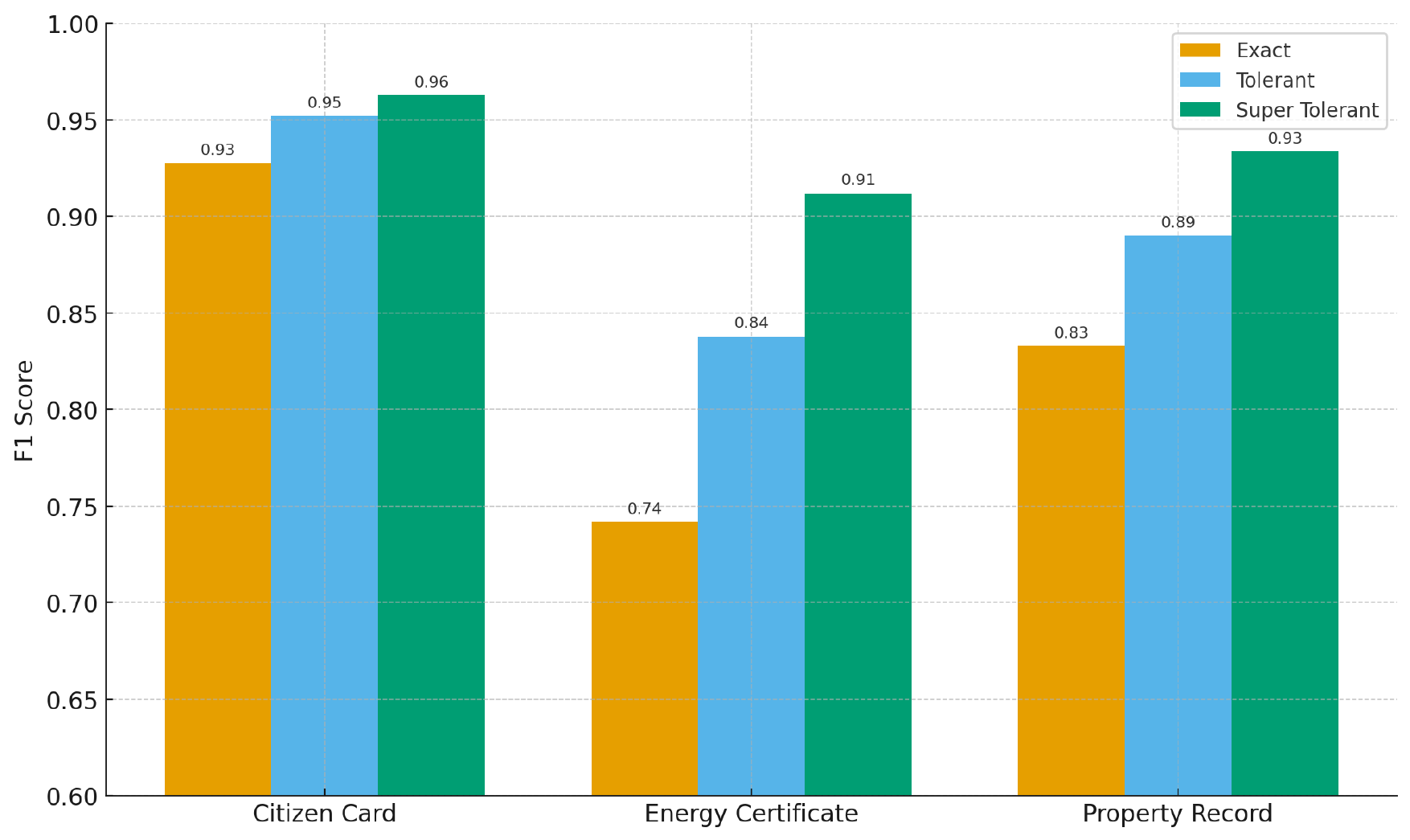}
    \caption{End-to-end pipeline F1-scores across matching tolerances}
    \label{fig:end-to-end-f1}
\end{figure}

Processing time per document remained practical: on GPU, average latency was 4.14s for Citizen Card, 8.25s for Energy Certificate, and 10.77s for Property Record.

\subsection{Comparison with Human Benchmark}
\begin{figure}[tb]
    \centering
    \includegraphics[width=\columnwidth]{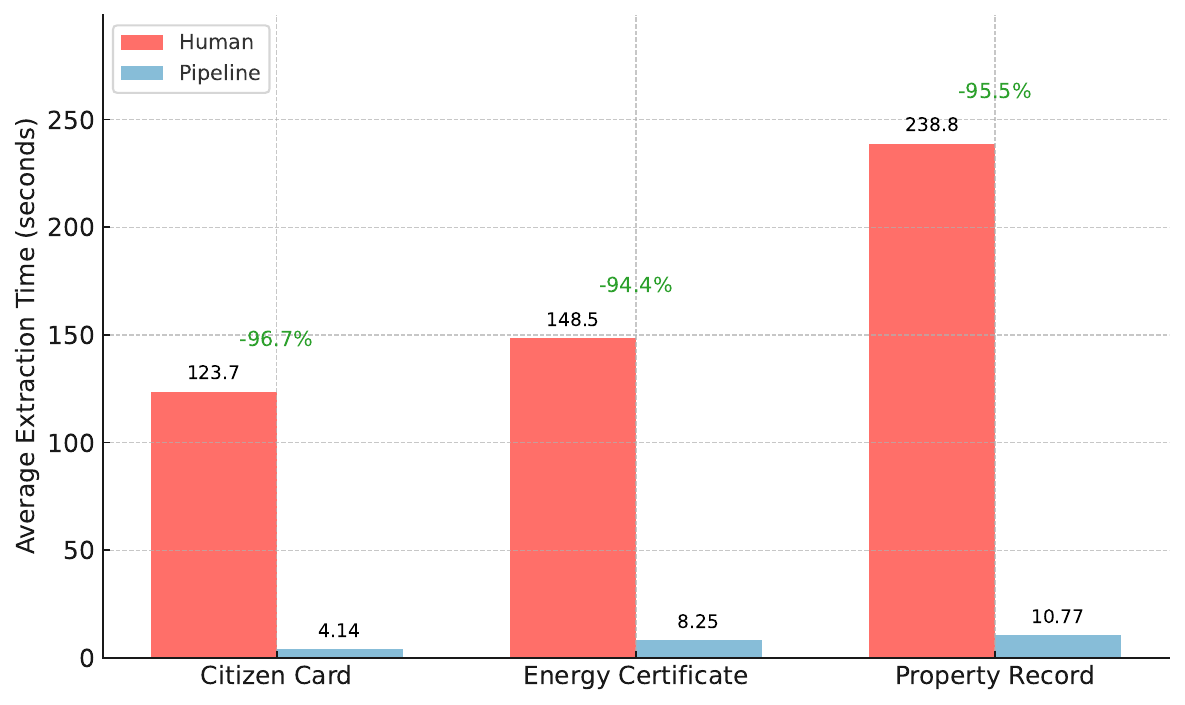}
    \caption{Average extraction time: human vs. pipeline}
    \label{fig:extraction_time_comparison}
\end{figure}

A total of nine document samples were annotated by multiple human evaluators to establish a performance baseline. The average human completion time ranged from 124 to 239 seconds per document, while the pipeline required only 4–11 seconds, representing a time reduction between 94\% and 97\%. Accuracy results show that the pipeline achieves near-human performance, with F1-scores differing by less than 3\% in tolerant and super-tolerant modes. As illustrated in Figure~\ref{fig:extraction_time_comparison}, automation provides substantial efficiency gains with minimal impact on accuracy.

\section{Conclusion}
\label{sec:conclusion}

This work addressed inefficiencies in real estate transactions by proposing the \textbf{Credentialing System}, a prototype that digitizes heterogeneous documents and transforms them into interoperable VCs. The system integrates an OCR and NLP pipeline for automated field extraction, a backend for credential life cycle management, and a user interface that supports Holder, Issuer, and Verifier roles. 

Evaluation in three types of documents demonstrated that synthetic datasets are effective in training high-performance extraction models, with LayoutLMv3 achieving over 0.99 F1 in held-out tests. The end-to-end pipeline achieved competitive accuracy under tolerant matching, with Citizen Card documents reaching 0.9628 F1, while maintaining practical throughput of under 11 seconds per document. A human benchmark confirmed that the pipeline operates an order of magnitude faster than manual annotation, approaching or even surpassing human accuracy under realistic conditions.

The main bottleneck remains OCR quality, with recognition errors propagating downstream and lowering strict match accuracy, particularly for visually complex documents such as Energy Certificates and Property Records. In addition, the prototype was evaluated primarily on synthetic data; performance on large-scale real-world corpora remains to be validated. The credential life cycle features, while functional, are simplified compared to production-grade deployments.


Overall, the results confirm that combining automated extraction with VC standards can substantially reduce manual effort, increase trust, and provide a scalable framework for secure digital transactions in the real estate sector and beyond.

This work was financially supported by the Project Blockchain.PT – Decentralize Portugal with Blockchain Agenda, (Project no 51), ref. C632734434-00467077, WP 6: Digital Assets Management, Call no 02/C05-i01.01/2022, funded by the Portuguese Recovery and Resilience Program (PRR), The Portuguese Republic and The European Union (EU) under the framework of Next Generation EU Program. This work was also supported by national funds, through Fundação para a Ciência e a Tecnologia I.P. (FCT) under projects UID/50021/2025 (DOI: https://doi.org/10.54499/UID/50021/2025) and UID/PRR/50021/2025 (DOI: https://doi.org/10.54499/UID/PRR/50021/2025).

\bibliographystyle{plain}
\bibliography{Bibliography.bib}

\end{document}